\documentclass{emulateapj}

\shorttitle{Dust properties at z=6.3}
\shortauthors{Stratta et al.}

\begin{document}

\title{Dust properties at z=6.3 in the host galaxy of GRB 050904}

\author{G. Stratta\altaffilmark{1}, R. Maiolino\altaffilmark{2},
    F. Fiore\altaffilmark{2} and V. D'Elia\altaffilmark{2}}
\altaffiltext{1}{ASI Science Data Center (ASDC), via G. Galilei,
   00040 Frascati, Italy}
\altaffiltext{2}{INAF -- Osservatorio Astronomico di Roma, via di Frascati 33,
    00040 Monte Porzio Catone, Italy}

\begin{abstract}

We investigate the dust extinction properties in the host galaxy of
GRB~050904 at z=6.29 by analyzing simultaneous broad band observations
of the optical and UV afterglow at three different epochs. We show
that the peculiar afterglow spectral energy distribution (SED)
observed at 0.5 days and at 1 day after the burst (1.6 and 3 hours
rest frame) cannot be explained with dust reddening with any of the
extinction curves observed at low redshift.  Yet,
the extinction curve recently inferred for the most distant BAL QSO at
z=6.2 nicely reproduces the SED of GRB~050904 at both epochs. Our
result provides an additional, independent indication that the
properties of dust evolve beyond z$\sim$6. We discuss the implications
of this finding within the context of the dust production mechanisms
through the cosmic ages. 

\end{abstract}



\section{Introduction}

Until recently, the investigation of the ISM at high redshift has been
mostly based on systems observed in absorption along the line of sight
of bright quasars or in Lyman break galaxies. However, the bright
emission of Gamma Ray Bursts (GRBs) and the broad redshift
distribution of these objects (from local to z=6.3, so far) has
highlighted GRBs as a new, independent tool to investigate the ISM at
high redshift.

GRB~050904 was a bright, long burst occurred at redshift z=6.29, the
most distant GRB identified so far.  Previous studies have shown that
the progenitor was a massive star embedded in a dense, metal enriched
molecular cloud \citep{campana06,frail06}.  This burst presents two
peculiarities with respect to lower redshift GRBs. First, X-ray data
at early times (within the first few minutes) shows
a large column of gas along the line of sight,
but little associated optical dust extinction \citep{campana06}.
Note also that the gas absorption is found to decrease rapidly within the
first few hours.
Second, its afterglow presents a peculiar flux suppression at $\rm
\lambda_{rest}\sim 1300~\AA$ (in addition to the Ly$\alpha$ blanketing
effect), at 0.5 days after the burst, while the flux redward 1600~\AA\ 
shows negligible extinction \citep{haislip06}.

In this work we further investigate these issues, and in particular
the latter, in terms of dust properties. We collect broad band
afterglow observations  (from near-IR to X-rays) at three epochs after the
burst, and fit the data with an intrinsic power-law and dust
reddening. In addition to the 'standard' extinction curves typically
assumed to model dust absorption in extragalactic objects, and in
lower redshift GRB's \citep{stratta04,stratta05,kann06}, we also
tested the extinction curve inferred from the most distant Broad
Absorption Line (BAL) QSO, at z=6.2 \citep{maiolino04}.  We show that
the latter extinction curve is in excellent agreement with the data of
GRB~050904, providing further, independent evidence for an evolution
of the dust properties at z$>$6.  We discuss the implications of our
results also for what concerns the origin of dust at high redshift.

\section{Multiband Spectral Analysis}

\subsection{X-rays and intrinsic SED}

The X-ray spectra have been extracted in the 0.3-10.0 keV energy range
from Swift X-ray Telescope (XRT) data following standard procedures.
At $\sim0.5$ days after the burst trigger $\rm t_0$ (2005, Sept. 5.07785 UT), 
the X-ray emission
shows a strong flaring activity \citep{cusumano07,gendre06}. In order
to avoid any possible spectral contamination from the flares in our
continuum analysis, we did not take into account data in temporal
intervals where flares were present. We fit the 0.3-10.0~keV 
data by assuming an absorbed power law. We find evidence for little absorption
in excess to the Galactic value
\citep[$\rm N_H^{Gal} = 5\times10^{20}~cm^{-2}$,][]{dl90},
in agreement with the analysis at late times by \cite{campana06}.
The best fit spectral energy index is $\alpha=1.2\pm0.2$ ($F_{\nu}\propto \nu^{-\alpha}$).
This is significantly steeper than the spectral index measured just after the
GRB event, suggesting that the synchrotron spectral break shifted toward lower
energies between the first XRT observation, $\sim3$~minutes after the GRB event, and
the observations at 12 hours \citep[in agreement with][]{cusumano07}.

The X-ray emission after 0.5 days decreases rather sharply, preventing
a detailed spectral analysis after this epoch. Thus, we are able to
directly compare the optical fluxes with the X-ray spectrum only at
0.5 days after the burst trigger.  However, the spectral and temporal
properties of this afterglow indicate that the cooling frequency is
redward of the optical range already few hours after the burst 
\citep{kann07,frail06,cusumano07}. As a consequence, no spectral break
is expected between the X-rays and the optical energy range at any of
the epochs of our analysis, nor the power-law index is expected to
change in such a time interval. 
Therefore, we can safely assume that the intrinsic
GRB spectrum at 0.5 days (observer frame) is a single power-law
extending from the X-rays to the optical.  At 1 and 3 days we assume
that the intrinsic SED has the same power-law index as at 0.5 days.

\subsection{Optical-NIR}

In order to build a broad band SED and investigate its possible
temporal variations, we collected multiband optical and near-IR
photometry at three epochs: $t_0+$0.5, 1 and 3 days after the burst
trigger $t_0$ (observer frame).  These specific epochs were chosen
because both near-IR and Z-band photometry (which are crucial to
investigate the rest-frame UV SED) were obtained around these times.
More specifically, at $\sim$0.5 days after the burst we used Z, J, H
photometry obtained with UKIRT and K' photometry obtained with IRTF 
\citep{haislip06}.  At $\sim$1 day after the burst we used J, H,
Ks, and z band photometry obtained with ESO/VLT by
\cite{tagliaferri05}.  At $\sim$3 days after the burst we used J, H,
Ks photometry from \cite{tagliaferri05} and z' obtained with Gemini
South by \cite{haislip06}.

Small temporal differences between the times of each individual
observation for the three selected  epochs were accounted for
by assuming that the optical-UV afterglow declines with time as a
broken power law. More specifically, the best fit indices to the light
curve obtained by \cite{tagliaferri05} are $\delta_1=-0.72\pm0.17$,
$\delta_2=-2.4\pm0.4$ and a break at $t_b=2.6\pm1$ days after the
burst (see also \cite{kann07}).

We corrected the observed magnitudes for the Galactic extinction toward the
direction of this burst (E(B-V)= 0.066) by assuming
the \cite{cardelli89} extinction curve with $R_V=A_V/E(B-V)=3.1$.

Fluxes in the z band were corrected for IGM Ly$\alpha$ blanketing effect
by  renormalizing the afterglow spectrum obtained by \cite{kawai06} (at 3~days)
so that its convolution with the specific z filter profile (including the
detector response) yields the photometry observed at each epoch. From the
re-normalized spectrum we derived the continuum flux density at 9300\AA \ (i.e.
just redward of the damped wing of the IGM Ly$\alpha$ absorption),
corresponding to $\rm \lambda _{rest} = 1275\AA$. As we shall see
in Sect.~\ref{dust_ev}, there is only little and marginal evidence
for any variation of the (rest-frame) UV SED between the epoch of the
\cite{kawai06} spectrum (3 days) and the previous epochs. This justfies the
extension of the IGM Ly$\alpha$ blanketing correction obtained at 3~days
to the other epochs.
Note that the z-band photometry in the three epochs is obtained with quite 
different filters at different telescopes. In particular, the z filter
(FORS2 at VLT) used by  
\cite{tagliaferri05} at 1 day after the burst is much redder 
than the z-band filters used by \cite{haislip06} at the other two
epochs. As a consequence, the z-band
correction factors for Ly$\alpha$ blanketing effect
are significantly different at the three epochs, and specifically: 
a factor of 3.02 at 0.5 days, 1.27 at 1 day (i.e. nearly
unaffected by IGM absorption), and 2.20 at 3 days.

\begin{figure*}[!]
\includegraphics[bb=30 160 585 610, angle=0,scale=.9]{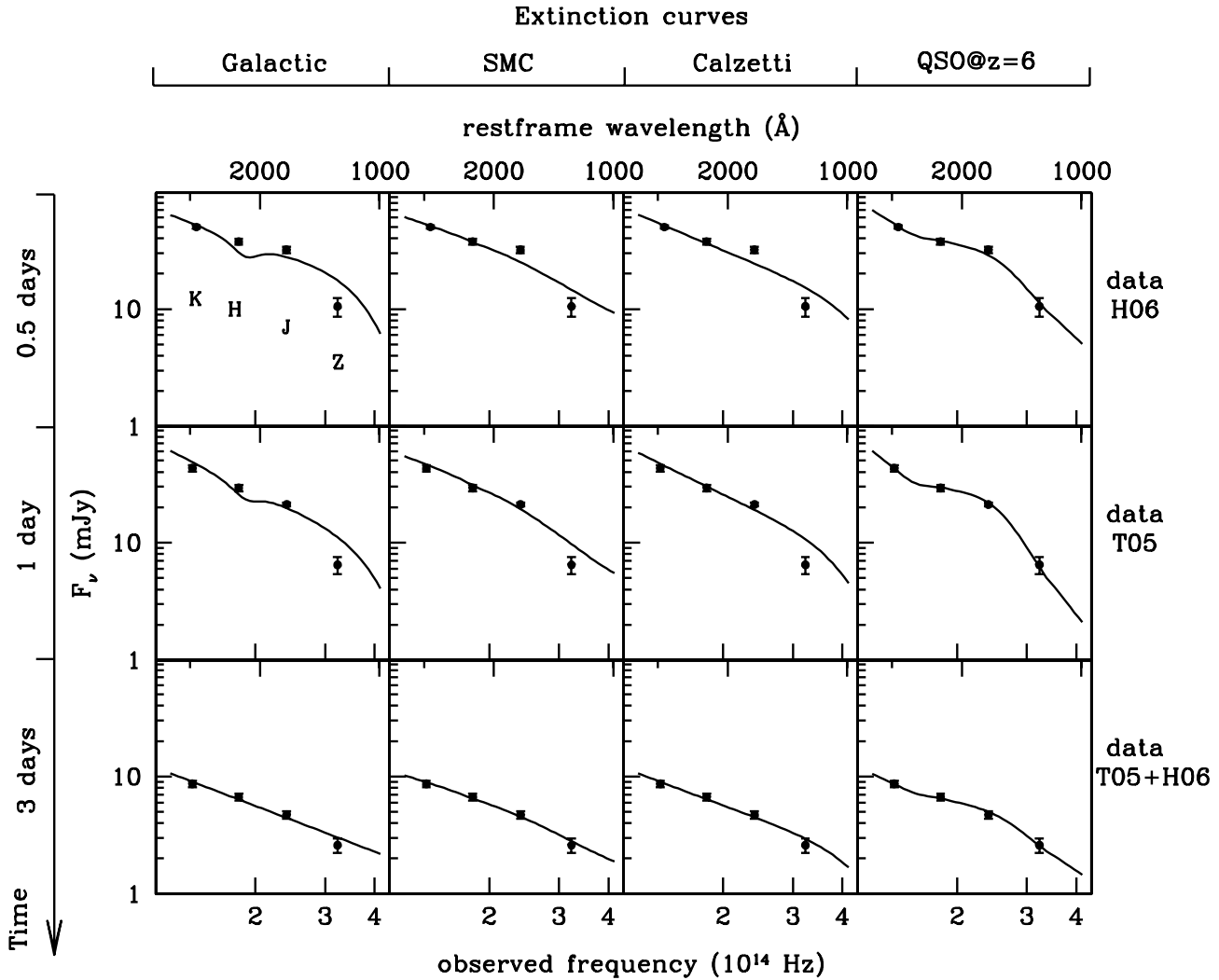}
\caption{optical-UV rest-frame SED of GRB~050904 at 0.5 (top), 1
(middle) and 3 (bottom) days after the burst (observer frame). The
corresponding observed bands are marked on the top-leftmost panel. The
source of data at each epoch are reported on the righ hand side (H06
= Haislip et al. 2006, T05 = Tagliaferri et al. 2005).  At each epoch
we show the best fit with each the four extinction curves shown in
Fig.~\ref{fig_ext}, as indicated on top of of each column.  }
\label{fig_sed}
\end{figure*}

\begin{table}
\caption{Best fit parameters of the GRB~050904 SED at different epochs.}
\centering
\begin{tabular}{lcc}
\hline
\hline
Extinction & $\rm A(3000\AA)^a$ & $\rm \chi^2/d.o.f.^b$  \\
curve      & mag   &                         \\
\hline
\hline
$t-t_0=0.5$ day    &    &     \\
\hline
  Galactic      & $0.53\pm0.08$         & 38.7/(12-2) \\
  SMC           & $0.33\pm0.06$         & 33.3/(12-2) \\
  Calzetti      & $0.43\pm0.06$         & 31.4/(12-2)  \\
 QSO@z=6       & $0.89\pm0.16$         & 15.1/(12-2)  \\
\hline
\hline
$t-t_0=1.0$ day    &  &   \\
\hline
  Galactic      & $0.43\pm0.18$   & 27.9/(4-2) \\
 SMC            & $0.48\pm0.08$   & 17.4/(4-2)  \\
  Calzetti      & $0.70\pm0.06$   & 24.3/(4-2)   \\
 QSO@z=6        & $1.33\pm0.29$   & 0.01/(4-2)  \\
\hline
\hline
$t-t_0$=3.0 day &  & \\
\hline
  Galactic      & $<0.3(2\sigma)$   & 3.2/(4-2) \\
 SMC            & $0.27\pm0.07$   & 1.1/(4-2)  \\
  Calzetti      & $0.40\pm0.07$   & 2.2/(4-2)   \\
 QSO@z=6        & $0.46\pm0.28$   & 0.6/(4-2)  \\
\hline
\hline
\end{tabular}
\tablenotetext{a}{Extinction at 3000~\AA. Errors are at $1\sigma$ confidence level.} 
\tablenotetext{b}{d.o.f.= degrees of freedom. Note that at 0.5 days after the
burst, X-ray data were included in the fit (see Section 2.1) thus yielding
a larger value of d.o.f.} 
\label{tab_fit}
\end{table}

\begin{figure}
\includegraphics[bb=20 145 590 715,angle=0,scale=0.4]{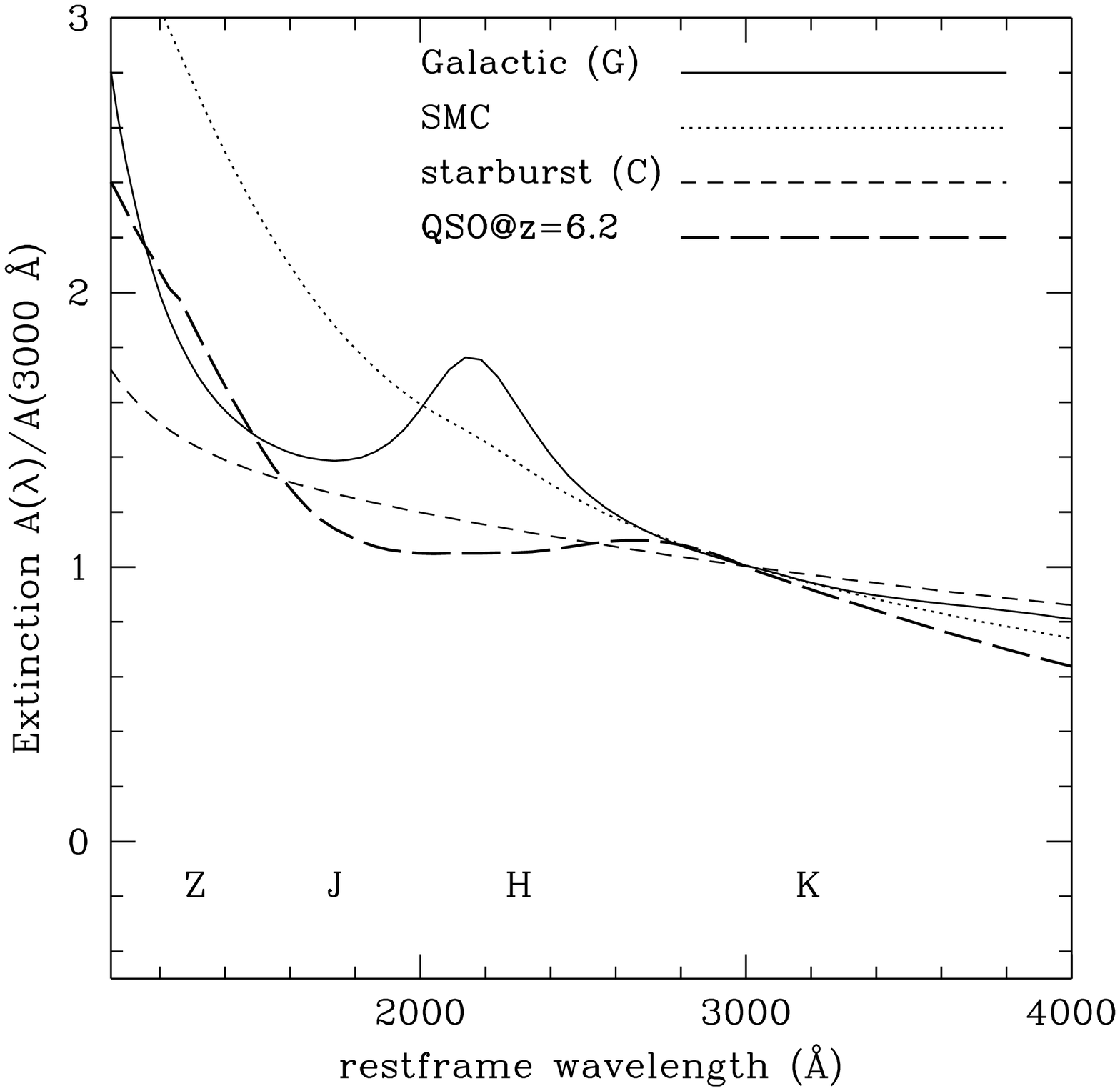}
\caption{Extinction curves tested in our analysis, normalized to
$\rm \lambda_{rest}=3000\AA $.
The photometric bands are shown shifted to the corresponding
central wavelengths at the rest frame of GRB~050904 at z=6.29.
The extinction curve inferred for the QSO at z=6.2 \citep{maiolino04}
is directly observed at $\rm 1260\AA < \lambda_{rest} < 3300\AA$, while
it is extrapolated beyond these limits by using the theoretical SN dust
extinction curve. However the latter extrapolation is unimportant
since for GRB~050904 at z=6.29 we essentially use the same
observed bands as for the QSO at z=6.2.}
\label{fig_ext}
\end{figure}

\section{Dust properties}

\subsection{Evidence for dust evolution at z$>$6}\label{dust_ev}

At 0.5 days after the burst, a peculiar flux suppression in the z band
($\lambda_{rest}\approx 1275\AA$) has been detected by
\cite{haislip06}, after accounting for IGM Ly$\alpha$ absorption.
We find the same z band flux suppression also at 1 day after the burst, with
a totally independent set of data (Fig.~\ref{fig_sed}).  The flux drop
is so sharp that \cite{haislip06} ruled out dust reddening 
as a possible cause.

We further investigate the dust extinction scenario to tackle the
issue of the peculiar z band flux suppression.  Beside the standard
extinction curves inferred in the local universe \citep[Galactic, SMC
and starburst attenuation curve,
from][]{cardelli89,pei92,calzetti94}\footnote{In the case of the starburst
attenuation curve we use the \cite{calzetti94} expression for the UV continuum
that, at 3000\AA, is 0.4 times the extinction for the ionized gas.}, 
we also consider the extinction curves computed by 
\cite{maiolino01} based on a dust model skewed toward
large grains, which apply to some low-z GRBs \citep{stratta04,stratta05},
and the extinction curve inferred by \cite{maiolino04} for the most distant
BAL QSO at z=6.19 (hereafter ``QSO@z=6'' extinction curve), illustrated in
Fig.~\ref{fig_ext}.

From the optical-to-X-ray spectral analysis, we find that at
$t_0+$0.5 days the optical data require additional extinction in
addition to the Galactic value to reproduce the flux values expected
from a power-law with the same spectral index of the X-ray spectrum
(see previous section). No 'standard' extinction curve can reproduce
the observed SED with the peculiar flux suppression in the z band, as
shown in the first three panels on the top of Fig.~\ref{fig_sed}. The
extinction curves from a dust model skewed toward larger grains also
fail to reproduce the flux in the z band, although they provide a
better fit to the J,H and K photometry than the 'standard' extinction
curves. The only extinction curve that provides an acceptable fit is
the QSO@z=6 one with $\rm A(3000\AA)\sim1$ mag (Fig.~\ref{fig_sed},
rightmost top panel).  The same result is obtained by using the
independent set of data at $t_0+$1 day (Fig.~\ref{fig_sed}, second
row of panels).

The statistics of the fits with the different extinction curves (along
with the inferred absolute extinctions) are given in
Table~\ref{tab_fit}. It is clear that the improvement of the fit with
the QSO@z=6 extinction curve is highly significant both at 0.5 and 1
days after the burst.

This result is still true at 3 days, though with a lower significance.
We also note that at the latter epoch
there is marginal evidence for a decrease of extinction (Tab.~\ref{tab_fit}).
This could be regarded as an indication of dust destruction by the blast
wave three days after the burst, or alternatively that at this late epoch
the fireball has a projected
size larger than the obscuring cloud and therefore the emitting region is less
absorbed. However, since the decrease of extinction at 3 days is only marginally
significant, we do not discuss this issue further and focus only on the
extinction curve properties.

Our result indicates that two totally different classes of objects at
$z\ge6$, QSOs and GRBs, are characterized by the same extinction
curve, which is different from that observed at lower redshift. More
specifically, the QSO at z=6.2 investigated by \cite{maiolino04} and
the GRB at redshift 6.3 investigated in this paper, provide
independent evidences for an evolution of the dust properties beyond
$z\sim 6$.

\subsection{The nature of dust at z$>$6}

A transition in the properties of the dust at $\rm z\sim 5-6$ is
theoretically expected.  Indeed, the major source of dust in the local
universe are the envelopes of AGB stars, which require about 1~Gyr to
evolve in large numbers \citep{dwek05,morgan03,marchenko06,todini01}.
At z$>$5 the age of the Universe is less than 1~Gyr, therefore AGB
stars fall short of time to produce enough dust.  As a consequence,
galaxies at z$>$5 should be devoid of most of the dust, since the
major source of dust (AGB) is missing.  Nonetheless, significant
masses of dust are still observed in distant QSOs up to z=6.4
\citep[e.g.][]{bertoldi03,priddey03,robson04,beelen06}.  Theoretical models have shown
that an alternative source of dust on short time scales are
core-collapse supernovae, which could therefore be the primary dust
production mechanism in the early universe
\citep{todini01,nozawa03,dwek05,morgan03,marchenko06}.  From the
observational point of view it is still not clear what is the actual
efficiency of dust production in SN ejecta
\citep{sugerman06,ercolano07,krause04,wilson05,green04}.  However,
both \cite{maiolino04} and \cite{hirashita05} found a good agreement
between the extinction curve expected from SN dust and the extinction
curve observed in the most distant BAL QSO at z=6.2 (QSO@z=6),
suggesting that in the latter object dust is predominantly produced by
SNe.  The result obtained in this paper indicate that also in the host
of GRB~050904 dust is likely produced mostly by SNe. Altogether, these
findings suggest that, more generally, regardless of the specific
class of object, at z$>$6 SNe are the main source of dust.

An independent evidence for different properties of the dust in the
host of GRB~050904 comes from the comparison between the large column
of absorbing gas observed at early times
\citep[N$_H\sim8\times10^{22}$ cm$^{-2}$,][]{cusumano07,campana06}
and the little
associated extinction observed in the optical-UV rest-frame.
More specifically, the latter implies an A$_V$/N$_H$ ratio more than
50 times lower than the Galactic value, which is unprecedented in lower
redshift GRBs \citep{stratta04,stratta05}.
\cite{campana06} ascribe this effect to dust mostly
composed of silicate grains (possibly produced by Pair Instability
SNe) which are destroyed soon after the burst. Here we
note that, independently of the dust composition, a strongly
reduced dust-to-gas ratio is naturally expected at z$\sim$6 as a
consequence of the dust evolutionary time scales. Indeed, as discussed
above, the lack of the important contribution to dust production from
AGB stars at z$>$6, makes the dust-to-gas ratio in the early universe
necessarily much lower than observed locally.

Furthermore, we note that the variation of dust content
as a function of redshift may be at the origin of the shortage of optical
GRB detections at lower redshift. Indeed, only
$\sim$50\% of the Swift GRBs have shown an optical counterpart, even by taking
advantage of the quick Swift GRB localization \citep{fiore07}.
The larger dust content at z$<$6, expected by the dust evolutionary scenarios,
may prevent the optical detection of low-z GRBs. This issue will be
discussed more extensively in a forthcoming paper.

\acknowledgments

We are thankful to the anonymous referee for useful coomments.
We are grateful to S.~Bianchi, R.~Schneider, A.~Ferrara and D.~Lazzati
for useful discussions. This work was supported by contracts 
ASI/I/R/039/04, ASI/I/R/023/05/0 and by the National Institute for Astrophysics
(INAF).

\end{document}